\documentclass[runningheads]{llncs}

\usepackage{graphicx}
\usepackage{todonotes}
\usepackage[T1]{fontenc}
\usepackage{float}

\begin{document}

\title{Participatory Action for Citizens' Engagement \\ to Develop \\ a Pro-Environmental Research Application}

\titlerunning{Participatory Action to Develop a Pro-Environmental Research App}

\author{Anna Jaskulska\inst{1,3}\orcidID{0000-0002-2539-3934} \and
Kinga Skorupska\inst{1,2}\orcidID{0000-0002-9005-0348} \and
Zuzanna Bubrowska\inst{1}\orcidID{ 0000-0003-1037-0630} \and
Kinga Kwiatkowska\inst{1}\orcidID{0000-0002-2957-3975} \and
Wiktor Stawski \inst{1}\orcidID{0000-0001-8950-195X} \and
Maciej Krzywicki\inst{1}\orcidID{0000-0002-8464-2830} \and
Monika Kornacka\inst{2}\orcidID{0000-0003-2737-9236} \and
Wiesław Kopeć \inst{1,2,3}\orcidID{0000-0001-9132-4171}}

\authorrunning{Jaskulska et al.}

\institute{Polish-Japanese Academy of Information Technology\\
\and
SWPS University of Social Sciences and Humanities \and
Kobo Association}

\maketitle              

\begin{abstract}
 To understand and begin to address the challenge of air pollution in Europe we conducted participatory research, art and design activities with the residents of one of the areas most affected by smog in Poland. The participatory research events, described in detail in this article, centered around the theme of ecology and served to design an application that would allow us to conduct field research on pro-environmental behaviours at a larger scale. As a result we developed a research application, rooted in local culture and history and place attachment, which makes use of gamification techniques. The application gathers air quality data from the densest network of air pollution sensors in Europe, thereby aligning the visible signs of pollution in the app with the local sensor data. At the same time it reinforces the users' pro-environmental habits and exposes them to educational messages about air quality and the environment. The data gathered with this application will validate the efficacy of this kind of an intervention in addressing residents' smog-causing behaviours.


\end{abstract}

\section{Introduction and Related Works}

Addressing the challenge of air pollution is a pressing matter for European governments, as despite increasingly better policies and green energy solutions Europe falls short of its zero pollution goal. In many places air pollution levels are above the alert thresholds set by WHO, which results in increased prevalence of health issues and higher mortality \cite{newsbbc}. In Poland, household heating is the source of almost half of PM10 and PM2.5 and 90$\%$ of PAH emissions. \cite{bilans2021}

To better understand this problem and come up with potential bottom-up solutions we conducted an extensive literature review and engaged in field research with residents of the area most affected by smog in Poland. The town of Myszków, is located in the Silesian Voivodeship, in Upper Warta River Depression. As a result of this location, smog generated by neighboring towns and villages settles in Myszków, which qualifies the town as one of the most polluted towns in Poland. \cite{smoglab_2020,administrator} The local authorities are aware of the problem of smog and have previously installed a couple of smog sensors, but despite such programs and efforts to subsidize clean energy solutions the problem persists. This awareness facilitated the town’s active participation in the VAPE ecological project conducted, among others, by the XR Lab of the Polish-Japanese Academy of Information Technology (PJAIT), the Institute of Psychology of the Polish Academy of Sciences (IP PAN), Warsaw University Faculty of Economic Sciences and Norwegian Institute For Air Research (NILU). The main goal of the VAPE project was to investigate how multi-sensory virtual experiences affect people's environmental behaviors. In this context we organized multiple participatory research, art and design activities with citizens. These were centered around the theme of ecology and aimed to co-develop an application that would allow us to conduct field research on how to encourage pro-environmental behaviours at a larger scale.

Based on our participatory design \cite{kopecspiral2018,KOPEC2019VirtuaATMParticipatory,KopecPDL2021}, citizen science \cite{zooni2021Interact} and living lab \cite{kopec2017living} experience, and an extensive literature review, especially in the area of environmental psychology and phenomenological geography as well as sociology we have developed an approach useful to engage citizens in pro-environmental projects by rooting these projects in the local environment, history and culture. This approach is backed by theories of place attachment \cite{LEWICKA2011207placeattachment} and place identity. Identification with a certain places can result in an increased quality of life and a greater commitment to protect one's habitat, neighbourhood and public spaces. This, in turn helps develop positive environmental behaviours. \cite{banka} We also made use of the concept of cognitive dissonance, which can be a powerful motivating force \cite{FestingerCognitiveDis1962}. We made use of this concept together with the ideas of both the "foot in the door" technique \cite{FootInTheDoor66}, and "nudges" \cite{ThalerSunstein08} as they can be helpful in shaping ecological behaviours. We were also inspired the role of children and young adults as environmental educators and ambassadors among their families. \cite{parentchild,parent2}

\section{Methods and Results}

\subsection{Participatory Research and Citizen Engagement}

Our activities with residents of Myszków were divided into three stages:

\begin{enumerate}
\item Participatory workshops at the local vineyard;
\item In-depth ethnographic research including digital ethnography and in-depth interviews with residents in their everyday contexts, city walks (both in the field and with the help of evaluative maps and questionnaires on multisensory perception of places important to the residents of Myszków);
\item Eco-picnic in the local activity park.
\end{enumerate}

Together with the residents we concentrated on the awareness of environmental issues and social causes of non-ecological behavior. We also focused on issues related to the residents' attachment to their city to find ways to engage them in pro-environmental activities. All of these events allowed the residents to open up to the issues raised by the VAPE project and engage in its activities.

\subsection{Participants}

Almost a hundred citizens of Myszków of all ages took part in our activities and research. Some of them were activists, involved in local formal and informal organizations and interest groups while others were just interested in what was happening in their city. Overall, we identified four types of residents in Myszków in terms of their attitudes towards the environment and air pollution: 

\begin{figure}[h]
 \centering 
 \includegraphics[width=\columnwidth]{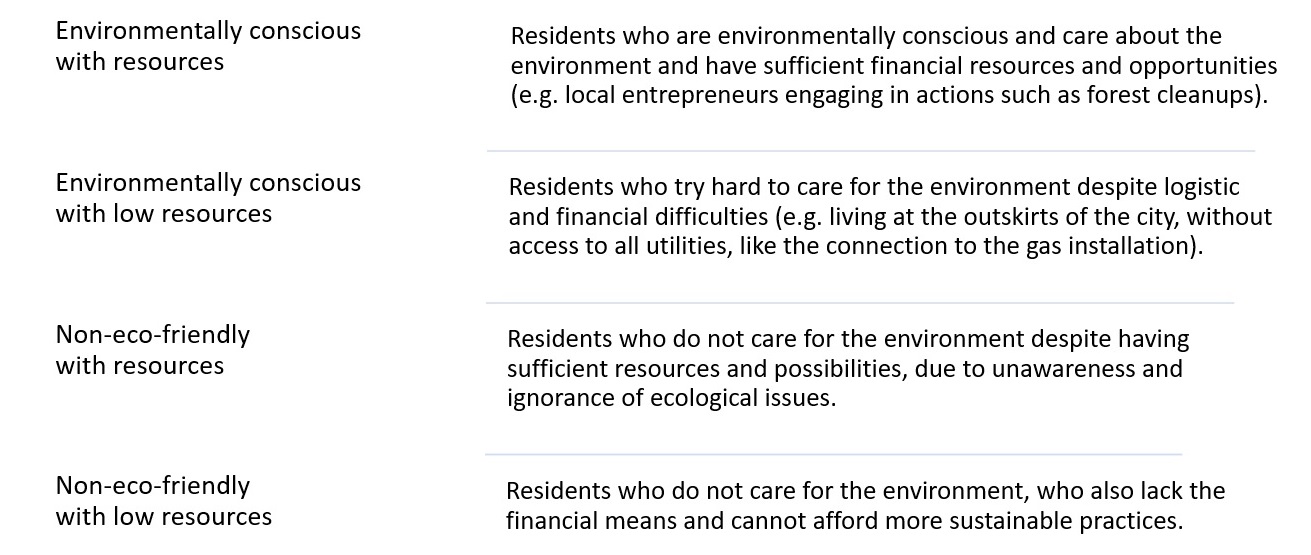}
\vspace{-5mm}
\end{figure}

\begin{figure}[h]
 \centering 
 \includegraphics[width=\columnwidth]{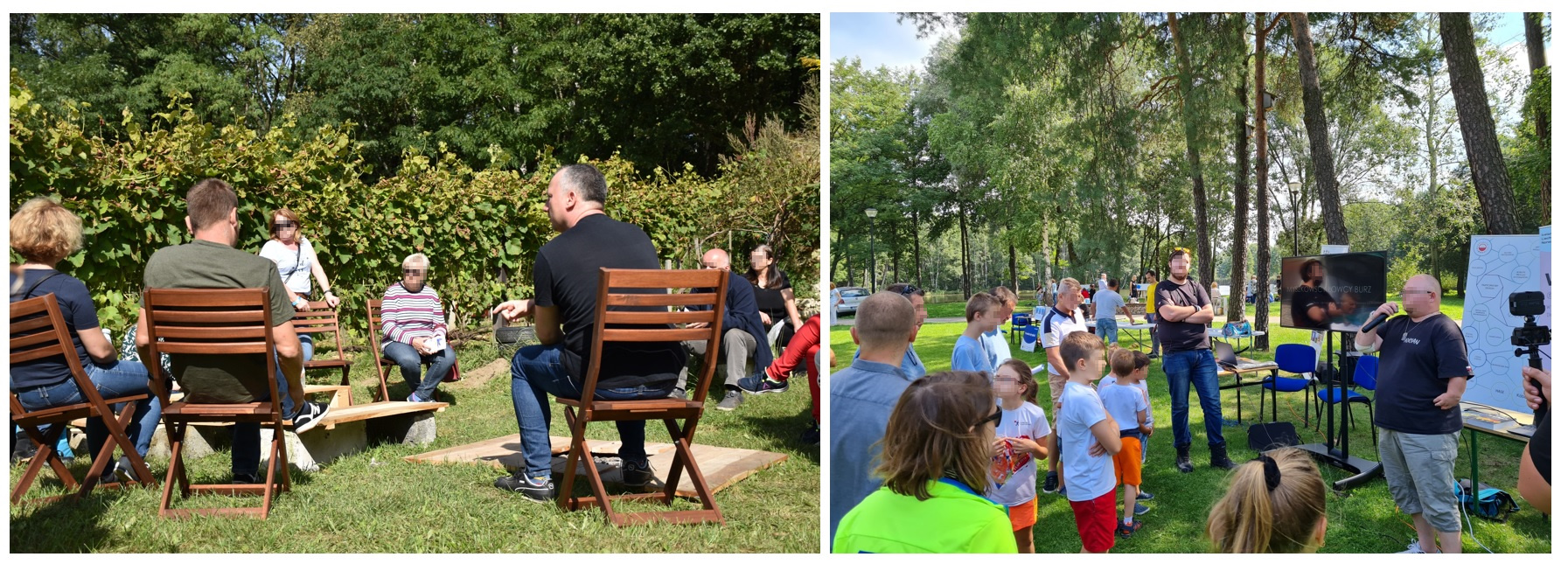}
 \caption{The outdoor locations: the vineyard and the activity park during project events}
 \label{fig:locations}
 \vspace{-3mm}
\end{figure}

\subsection{Locations}
Our participatory research activities with citizens took place directly in the field (see Fig. \ref{fig:locations}) which allowed us to eliminate the laboratory environment and workshop rooms. To organize our events we selected the only two "places of interest" in Myszków listed on TripAdvisor, a popular travel platform: a local vineyard and an activity park. Both provided outdoor facilities, which was important due to the COVID-19 pandemic, and were well-aligned with the green theme of our project. What follows is a detailed description of the scenarios of these activities as well as a summary of insights gained in their course.

\subsection{Participatory Workshops at the Local Vineyard}

\begin{figure}[h]
 \centering 
 \includegraphics[width=\columnwidth]{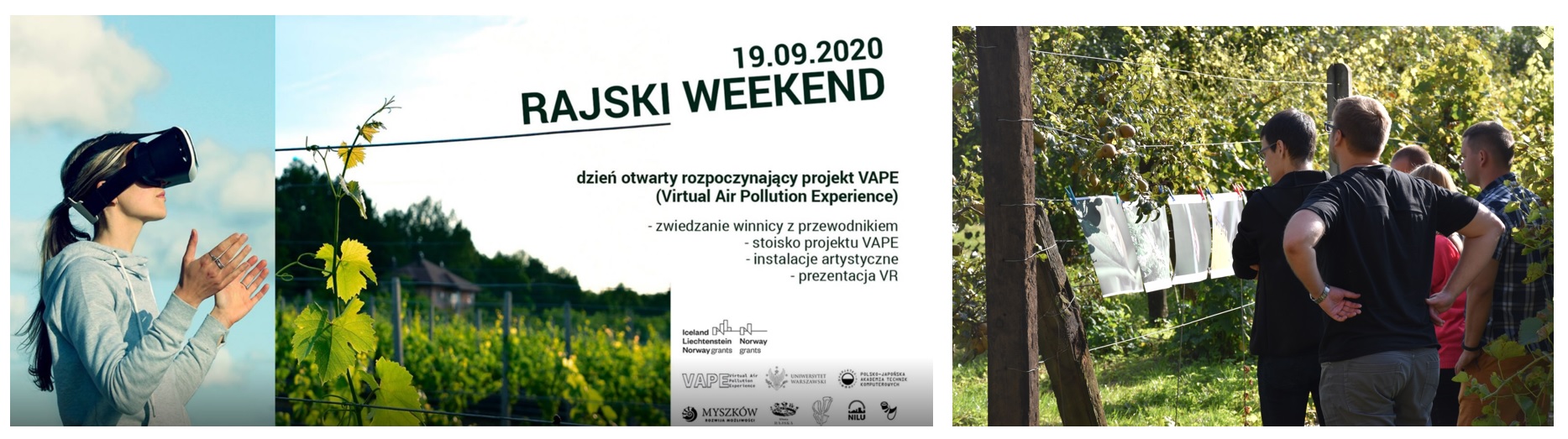}
 \caption{Guided tour of the vineyard and macro photography exhibition}
 \label{fig:rajski_weekend}
  \vspace{-3mm}
\end{figure}

The participatory processes at the vineyard were supported by activities related to air and water pollution, sustainability and environmental education as well as presentations of new technologies that can support these. Participants took part in a guided tour of the vineyard and an exhibition of macro-photographs of insects threatened with extinction by climate change (see Fig. \ref{fig:rajski_weekend}).

\begin{figure}[h]
 \centering
 \includegraphics[width=\columnwidth]{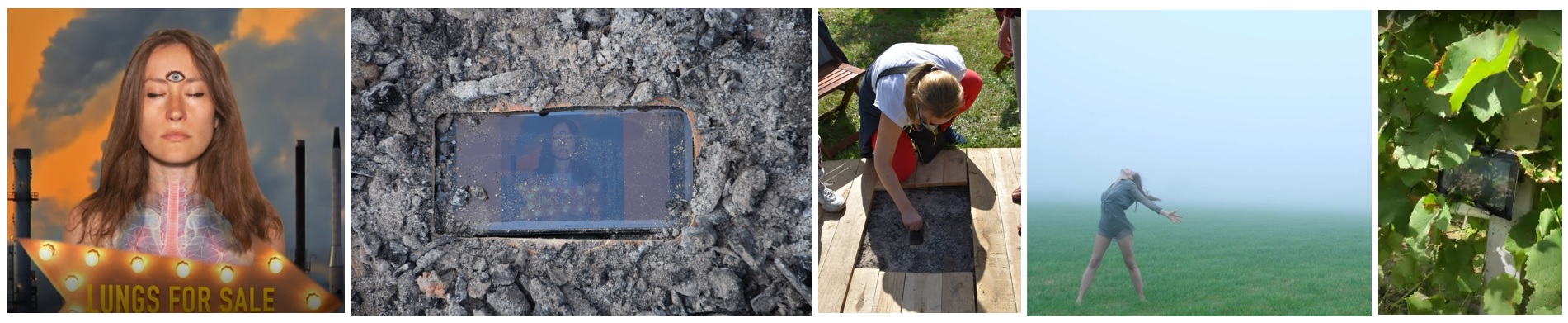}
 \caption{"Lungs for sale" and "Wake up", video artworks by Aleksandra Karpowicz}
 \label{fig:art}
  \vspace{-3mm}
\end{figure}

At the vineyard we also placed two artworks by Aleksandra Karpowicz, a London-based video artist. One of them, called "Lungs for sale", which depicts the coughing artist, was played on a loop on a smartphone buried in the ash from a campfire. The visitors, tired of coughing, could walk into a row of vines where, in the second video, Alexandra is dancing in the meadow, and her message "wake up" can be heard in the background (see Fig \ref{fig:art}).

Aleksandra Karpowicz also prepared a video manifesto in which she addressed issues related to climate change, air and earth pollution, showing them from the social, political and economic points of view. We built a wooden TV set stand in order to fit it into the natural surroundings where the video was played. These works of art served the same purpose as a lecture would: they introduced the participants to the topic of the workshop. We used them to provoke discussions with various groups of residents, who were more open to share their thoughts because the artist did it first. 
The second empowerment element consisted of demonstrations of solutions related to virtual, augmented and mixed reality. During our workshop with the participants we brainstormed, facilitated discussions and collected insights.

\subsection{In-depth Interviews and Ethnographic Research}

\begin{figure}[h]
 \centering
 \includegraphics[width=\columnwidth]{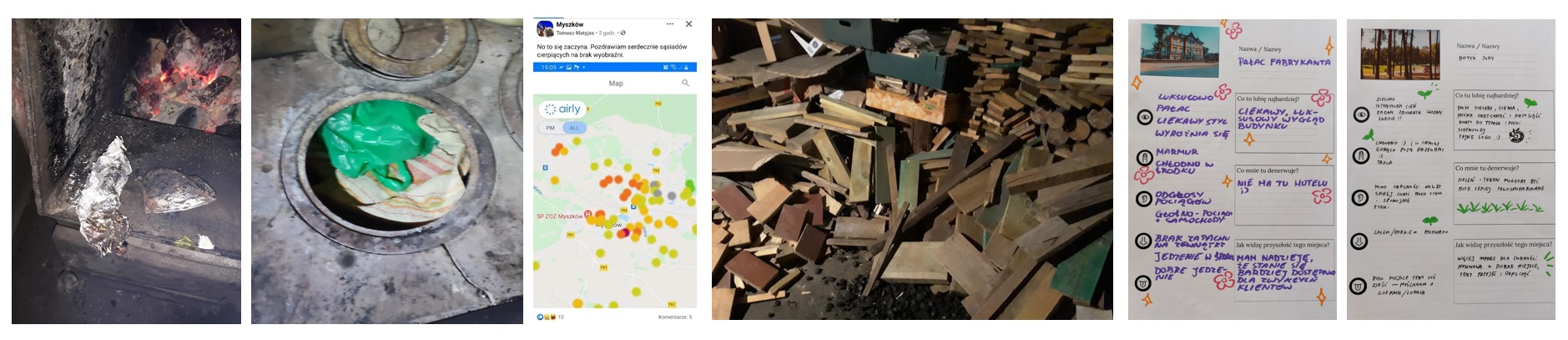}
 \caption{Examples of Digital Ethnography research from Myszków's social media and cards from city walks with photos printed with a portable photo printer}
 \label{fig:digital_etnography}
  \vspace{-4mm}
\end{figure}

In order to deepen the insights gathered during the participatory workshops, we conducted qualitative research including 
individual and group in-depth interviews; digital ethnography - comprehensive research of local websites and social networks, like public and private groups for residents, profiles of the town, municipal police, social and cultural organizations, archives of town history, etc.; ethnographic walks (including a "virtual one" using only the map) combined with multidimensional descriptions of selected places; exploratory walks of people who have never been to Myszków - using urban signposting and data from geolocation-based games and travel applications (see Fig. \ref{fig:digital_etnography}). Additionally, a series of quantitative studies in the form of online surveys are underway.

\subsection{Family Ecological Picnic}

\begin{figure}[h]
\centering 
\includegraphics[width=\textwidth]{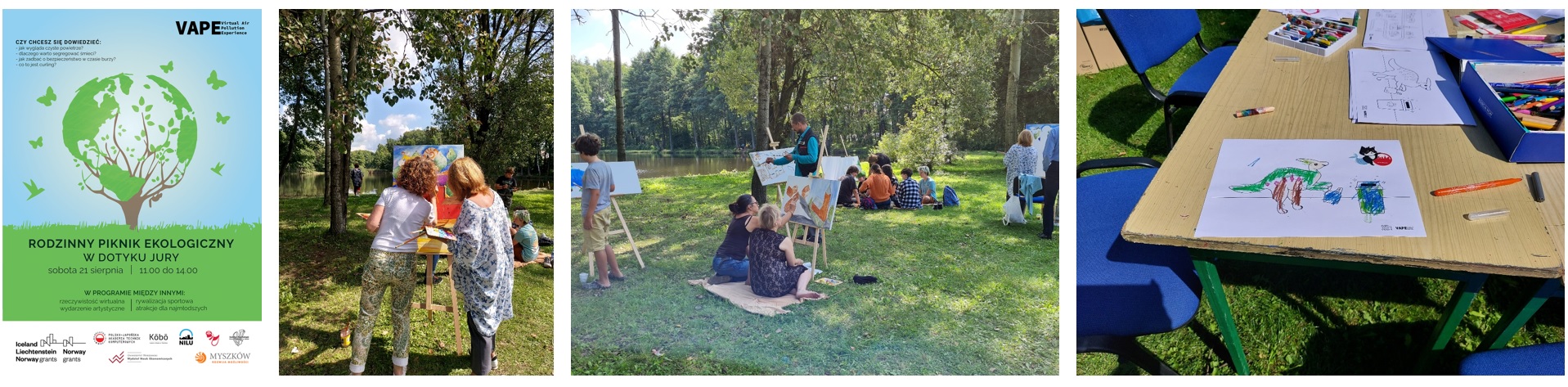}
\caption{Artistic activities with residents of all ages during the eco-picnic}
 \label{fig:plener}
   \vspace{-3mm}
\end{figure}

During the eco-picnic together with the inhabitants of Myszków we conducted activities directly involving residents. Adults could participate in the creative process to paint Myszków as the city of their dreams and children played with colouring books on waste segregation. (see Fig. \ref{fig:plener}).  

\begin{figure}[h]
\centering 
\includegraphics[width=\textwidth]{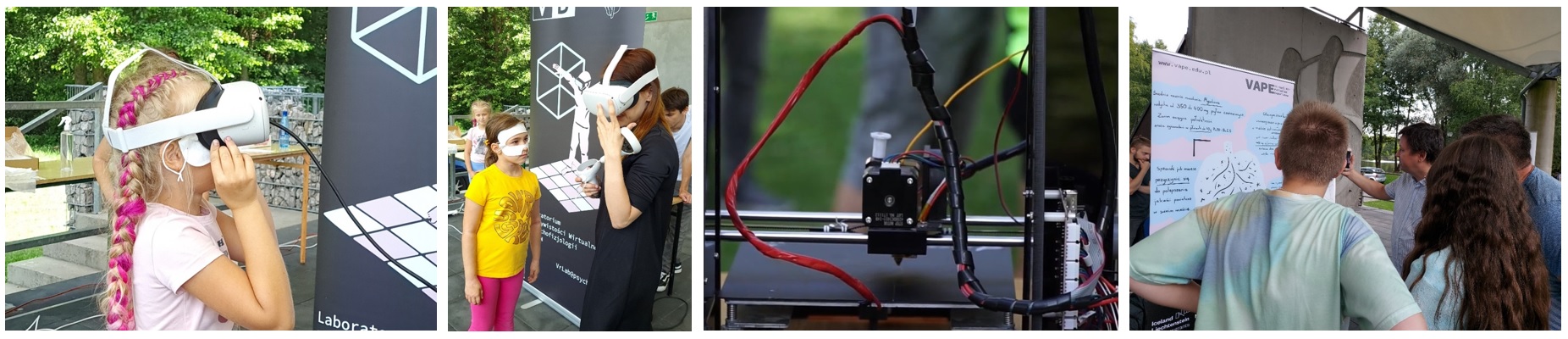}
\caption{VR, AR \& 3D printing during eco-picnic}
 \label{fig:VR}
   \vspace{-5mm}
\end{figure}

We gave the citizens of Myszków the possibility to test our project solutions in VR, AR and mobile apps. We also showcased 3D printers used in the VAPE project to prepare objects for upcoming research on the perception of touch in virtual reality (see Fig. \ref{fig:VR}). We invited a talk on violent weather phenomena and organized sports activities, related to the seasons, such as beach volleyball or curling (presented as a board game).

\section {Discussion}

In the course of our research we identified several key areas of non-ecological behavior among residents that contribute to smog formation. These were mostly  related to heating their homes and included lack of access to municipal heating, no access to gas infrastructure in the neighbourhood, and thus gas heating, the use of old stoves without adequate filters, the use of improper techniques of burning in the stove (increasing the amount of smoke) and burning garbage. Some of them are determined by administrative reasons, for example, receiving EU subsidies to replace the stove with a more ecological one is possible only after the specific house has been modernized using the interested person's own funds. In some cases the problem is caused by poverty and lack of funds to buy appropriate heating fuel. We have excluded these factors from our area of focus as they are related to the environmental and social policies, and can not change as a result of our project. The city authorities are aware of these problems and try to actively counteract them, e.g. by creating their own programs for stove replacement without additional conditions. 

Another non-ecological activity in Myszków is using garbage to heat homes: household litter, used tires or old painted furniture. The scale of the problem is illustrated by the reports from the municipal police interventions. There are even cases of burning garbage in new stoves designed for ecological fuel. The problem persists even though waste collection fees do not depend on quantity, so throwing garbage into the stove does not result in direct savings. Causes of litter burning, according to our participatory research include: lack of knowledge about the consequences of burning garbage, viewing it as a way to get rid of garbage as quickly as possible, e.g. burning leaves without waiting for them to be collected, lack of knowledge about the possibility to dispose of difficult garbage such as used tires, old furniture, renovation waste, free of charge at the local selective waste collection point, viewing garbage as valuable fuel, especially in the cooler seasons of autumn and spring. In the course of further analysis we decided that trying to influence this behavior through instilling proper garbage disposal and sorting, as well as choosing the right fuel for heating, ought to become a key issue addressed in our mobile field research application.

\section {Mobile Field Research Application}

The Vape Mobile app combines a Match-3 game concept (inspired by Bejeweled, Candy Crush) with a simulation of how the air quality in Myszków affects both the town and the daily life of its inhabitants. Based on the "foot in the door" technique, the garbage sorting and choosing appropriate heating sources in the game is meant to be the first step to feeling environmentally engaged and making pro-environmental choices in the real world.

\begin{figure}[h]
 \centering
 \includegraphics[width=\columnwidth]{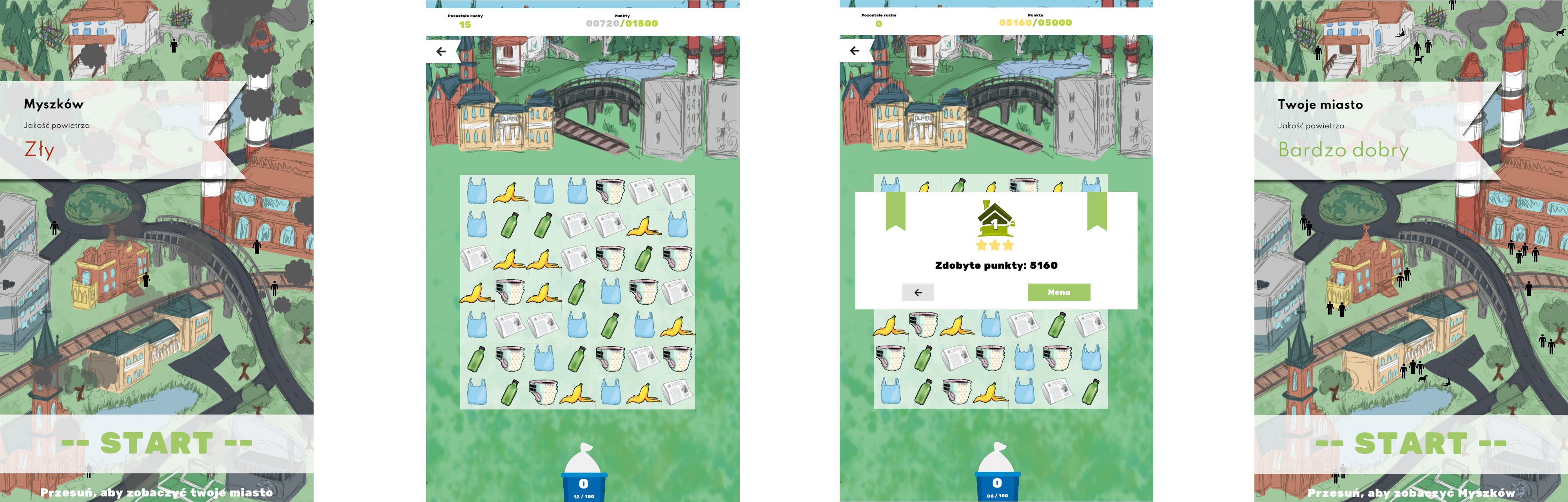}
 \caption{Screens from the pro-environmental research application to address the problem of smog.From left to right: 1) The user sees that the local air quality is not very good 2) they can engage with the game where they sort trash, or choose better energy sources 3) their performance is saved 4) if air quality improves, then the application skin changes }
 \label{fig:app}
   \vspace{-3mm}
\end{figure}

The application was designed to have simple mechanics and be fit for people of different ages, including children and older adults. The application was to be set in the local context - the player improves their eco-Myszków and progress is achieved by sorting garbage and choosing appropriate fuels for heating (see Fig. \ref{fig:app}). Gamification elements such as educational quizzes, quests or leaderboards ought to encourage regular use without being time-consuming. The game also ought to provide exposure to smog visualizations to continue laboratory studies in the field. By design the participant, in order to gain additional points in the game, in the frequency determined by the researchers, is exposed to visualizations of pollution or clean air. The user can also compare the state of ‘their Myszków’ and the real life one, based on the actual readings from air pollution sensors (this functionality uses the sensor network in Myszków).

The app was created with the use of Unity for Android and common algorithms used in Match-3 type games.

All assets used in the game are either free to use or were created by us using Procreate or Clip Studio Paint. During our research we collected references to properly portray Myszków and make sure it is recognizable in the app at first glance. Many landmarks such as the historical railway station, a local palace and the bandshell from activity park were used to achieve this. The information on recycling trash is based on regulations valid in Myszków.

\section{Conclusions and Further Work}

Thanks to the participatory research and art activities we were able to create a field research application for raising awareness about smog. Based on our experience, to engage in similar project activities, we can recommend working closely with local authorities, institutions, community organizations and activists as well as eco-friendly stakeholders. Digital ethnography, including a review of local social networks, is also worth including in this type of research. In the next steps we will continue to develop and test new application functionalities. After completing the first cycle of quantitative remote survey-based research we will be able to inform the application design with insights from the results of survey data analyses. Overall, empowerment and community inclusion in the research, design and development processes, for which both art and new technologies can be used are powerful vessels to discover real needs of potential users. Such activities ought to constitute a crucial part of impact-focused social change projects enabled by novel technologies.

\section*{Acknowledgments}

The VAPE project, which made this research possible, received funding from the IdeaLab competition for interdisciplinary research projects. It was funded by the European Economic Area Financial Mechanism 2014 - 2021 (grant number 2019/35/J/HS6/03166). We would also like to thank the many people and institutions gathered together by the distributed Living Lab Kobo and HASE Research Group (Human Aspects in Science and Engineering) for their support of this study.
\bibliographystyle{splncs04}
\bibliography{bibliography}

\end{document}